\documentstyle[12pt]{article}
\renewcommand{\today}{20 October 1995}
\makeatletter
\@ifundefined{selectfont}{
  \def\mathrm#1{{\rm #1}}
}{
  \let\oldmathrm=\mathrm
  \def\mathrm#1{{\oldmathrm{#1}}}
}
\@ifundefined{reset@font}{\let\reset@font\empty}{}
\def\d{\,\mathrm{d}}
\def\e{\mathop{\mathrm{e}}\nolimits}
\def\Bbb#1{{\bf\relax#1}}
\def\section{\@startsection {section}{1}{\z@}{-3.5ex plus-1ex minus
    -.2ex}{2.3ex plus.2ex}{\reset@font\normalsize\bf\boldmath}}
\def\subsection{\@startsection{subsection}{2}{\z@}{-3.25ex plus-1ex
    minus-.2ex}{1.5ex plus.2ex}{\reset@font\normalsize\it}}
\def\subsubsection{\@startsection{subsubsection}{3}{\z@}{-3.25ex plus
 -1ex minus-.2ex}{1.5ex plus.2ex}{\reset@font\normalsize\bf}}
\advance\textheight by \topskip
\makeatother

\def\SU{\mathord{\rm SU}}
\def\Pf{\mathord{\rm Pf}}

\def\tr{\mathop{\rm tr}\nolimits}

\def\id{{\bf 1}}
\begin{document}
\bibliographystyle{paper2}
%
%
\thispagestyle{empty}
\begin{flushright}
  hep-th/9510149\\DAMTP 95-52
\end{flushright}
\vskip 2em
\begin{center}\LARGE
  Curiosities at $c=-2$
\end{center}\vskip 1.5em
\begin{center}\large
  Horst G. Kausch\footnote{Email: {\tt H.G.KAUSCH@DAMTP.CAM.AC.UK}}
\end{center}
\begin{center}\it
Department of Applied Mathematics and Theoretical
Physics, \\
University of Cambridge, Silver Street, \\
Cambridge CB3 9EW, U.K.
\end{center}
\vskip 1em
\begin{center}
  \today
\end{center}
\vskip 1em
\begin{abstract}
  Conformal field theory at $c=-2$ provides the simplest example of a
  theory with ``logarithmic'' operators. We examine in detail the
  $(\xi,\eta)$ ghost system and Coulomb gas construction at $c=-2$ and
  show that, in contradistinction to minimal models, they can not be
  described in terms of conformal families of {\em primary\/} fields
  alone but necessarily contain reducible but indecomposable
  representations of the Virasoro algebra.  We then present a
  construction of ``logarithmic'' operators in terms of ``symplectic''
  fermions displaying a global $SL(2)$ symmetry.  Orbifolds with
  respect to finite subgroups of $SL(2)$ are reminiscent of the $ADE$
  classification of $c=1$ modular invariant partition functions, but
  are isolated models and not linked by massless flows.
\end{abstract}

\section{Introduction}
\label{sec:intro}

Two-dimensional conformally invariant field theories feature
prominently in two areas of theoretical physics: they provide the
perturbative vacua of string theory and they serve as a framework for
understanding second order phase transitions of statistical systems.
The success of conformal field theory is due to the fact that in two
dimensions conformal transformations coincide with analytic coordinate
transformations generated by the holomorphic and anti-holomorphic
components of the stress tensor.  This results in an
infinite-dimensional local conformal algebra consisting of two
commuting copies of the Virasoro algebra.  The conformal anomaly $c$,
called the central charge of the Virasoro algebra, is the main
parameter characterising a conformal field theory.  In their seminal
paper \cite{BPZ} Belavin, Polyakov and Zamolodchikov showed that for
the minimal series, $c=1-6(p-p')^2/pp'$ with coprime integers
$p,p'>1$, there are only a finite number of irreducible
representations of the Virasoro algebra.  In such a situation the
space of states decomposes into representations of the left and right
Virasoro algebra and the partition function can be written as a finite
sum over holomorphic times anti-holomorphic characters.  A conformal
field theory with periodic boundary conditions can naturally be
thought of as defined on a torus. This implies the invariance of the
partition function under modular transformations.  Using this
requirement Cappelli, Itzykson and Zuber were able to classify the
partition functions for the minimal models \cite{CIZu1,CIZu2}.

Going beyond the minimal series the simplest example of a conformal
field theory is provided by a free massless scalar field. The Virasoro
algebra generated from the stress-energy tensor has central charge
$c=1$.  If one compactifies the scalar field to take values on a
circle of radius $\rho$ the primary fields are classified by their
momenta and winding numbers which, by locality, are forced to lie on a
lattice resulting in the Coulomb gas partition function $Z(\rho)$.
 At the radius $\rho=1$ one obtains the $\SU(2)$
WZW model at level one which possesses a global $\SU(2)$ symmetry. In
this situation one can form orbifold models \cite{DVVV1} for any
subgroup $\Gamma\subset\SU(2)$ introducing additional sectors with
twisted boundary.  From the A-D-E classification of finite subgroups
of $\SU(2)$ \cite{Mcka1} one can obtain a classification of $c=1$
modular invariant partition functions \cite{Gins1,Harr1}.

Attempts to study more general conformal field theories beyond the
above examples have produced a wealth of examples but their
understanding is still rather sketchy. In this paper we shall consider
models at $c=-2$, in particular the fermionic $(\xi,\eta)$ system. It
is related to the continuum limit of dense polymers \cite{Sale1} and
is generated by two fermionic ghost fields of weight zero and one.  It
is also of interest for the ghost sector of superstring theory
\cite{FMSh1}.  The system has a $U(1)$ symmetry allowing the
introduction of twist fields.  The partition function of the
$(\xi,\eta)$ system with ${\Bbb Z}_{2N}$ twist is the same as that of
a free boson compactified on a circle of radius $N\sqrt2$, however
with a different choice of vacuum state.  The system is thus modular
invariant but has a charge asymmetry and non-vanishing correlation
functions require the insertion of a $\xi$ zero-mode.  The model also
contains fields which are neither primary fields nor descendants of a
primary field. They form reducible but indecomposable representations
of the Virasoro algebra.

The same features occur in the Coulomb gas construction of minimal
models where one introduces screening charges and performs a BRST
projection to remove the reducible representations.
In the case of the $(\xi,\eta)$ system the analogous procedure is to
go to the kernel of $\eta_0$. This defines a ``small'' algebra generated by
$\partial\xi$ and $\eta$. On the ``small'' algebra the $U(1)$ symmetry
is enhanced to an $SL(2)$ symmetry. This symmetry is only global and
hence not generated by a Kac-Moody algebra but related to a W-algebra.
Twisting the ``small'' algebra results in characters which
do not transform in any simple way under the modular group.

The 4pt function of the twist field calculated from the null vector
and monodromy constraints has a logarithmic singularity.  This implies
that the fusion of twist fields in the ``small'' algebra results in
fields forming two-dimensional Jordan cells for $L_0$ \cite{Gura1}.
These ``logarithmic'' operators extend the space of fields of the
``small'' algebra and we present a construction in terms of
``symplectic'' fermions displaying a $SL(2)$ symmetry.
We construct orbifolds with respect to finite subgroups of $SL(2)$,
following the procedure employed in the $c=1$ case \cite{Gins1}, and
investigate the effect of perturbing the orbifold models by marginal
operators.

\section{The $(\xi,\eta)$-ghost system}
\label{sec:ghost}

The $(\xi,\eta)$ system is defined by the action
\begin{equation}
  S = \frac1{2\pi} \int\!{\d}^2z \left( \eta\bar\partial\xi +
  \bar\eta\partial\bar\xi \right),
\end{equation}
where $\xi$ and $\eta$ denote holomorphic fermionic ghost fields of
dimension 0 and 1. The corresponding operator product expansion is
\begin{equation}\label{eq:xieta-ope}
  \xi(z) \eta(w) = \eta(z) \xi(w) = \frac{1}{z-w} + O(1).
\end{equation}
Equivalent relations are satisfied by anti-holomorphic ghost fields
$\bar\xi$ and $\bar\eta$. When considering chiral fields we shall
frequently omit mention of the anti-holomorphic components.
Introducing the mode expansion
\begin{equation}
  \xi(z) = \sum_{n\in\Bbb{Z}} \xi_n z^{-n}, \qquad
  \eta(z) = \sum_{n\in\Bbb{Z}} \eta_n z^{-n-1},
\end{equation}
we obtain from the short distance singularity (\ref{eq:xieta-ope}) the
anti-commutator
\begin{equation}\label{eq:xieta-acr}
  \{ \xi_m, \eta_n \} = \delta_{m+n,0},
\end{equation}
while the other anti-commutators vanish.
The stress tensor has central charge $c=-2$ and is given by
\begin{equation}\label{eq:Txieta}
  T(z) = \sum_{n\in\Bbb{Z}} L_n z^{-n-2} =
  \mathopen: \partial\xi(z) \eta(z) \mathclose:,
\end{equation}
where the $:\cdots:$ denotes fermionic normal ordering, that is, the
annihilation modes are moved to the
right with a minus sign whenever two fields are interchanged.
The $sl(2,\Bbb{C})$-invariant vacuum $\Omega$ is characterised by
$\xi_m \Omega = 0,\, \eta_n \Omega = 0$, for $m>0, n\geq0$.

The $(\xi,\eta)$ system has a $U(1)$ symmetry which is generated by
the natural $U(1)$ current
\begin{equation}\label{eq:U1-curr}
  J(z) = \mathopen: \xi(z)\eta(z) \mathclose:,
\end{equation}
which counts $\xi$ with charge $1$ and $\eta$ with charge $-1$.
We can thus consider twist fields $\sigma_\lambda$ such that
\begin{eqnarray}
  \xi(\e^{2\pi i}z) \sigma_\lambda(0) &=&
  \e^{2\pi i\lambda} \xi(z) \sigma_\lambda(0),
  \\
  \eta(\e^{2\pi i}z) \sigma_\lambda(0) &=&
  \e^{-2\pi i\lambda} \eta(z) \sigma_\lambda(0).
\end{eqnarray}
This implies that acting on the twisted sector ${\cal M}_\lambda$
generated by $\sigma_\lambda$
the fermion fields have boundary conditions in the angular direction
\begin{displaymath}
  \xi\mapsto\e^{2\pi i\lambda}\xi,
  \qquad \eta\mapsto\e^{-2\pi i\lambda}\eta.
\end{displaymath}
and a mode expansion
\begin{equation}
  \xi(z) = \sum_{m\in\Bbb{Z}} \xi_{m-\lambda} z^{-m+\lambda}, \qquad
  \eta(z) = \sum_{m\in\Bbb{Z}} \eta_{m+\lambda} z^{-m-1-\lambda}.
\end{equation}
The operator product expansion of $\xi(z)$ and $\eta(w)$ remains the
same when acting on a twisted representation.
By multiplying it with appropriate powers of
$z$ and $w$,
\begin{equation}
  z^{-\lambda} w^{\lambda} \xi(z) \eta(w) =
  z^{-\lambda} w^{\lambda} \left(
  \frac{1}{z-w} + J(w) + (z-w) T(w) + \cdots \right),
\end{equation}
we get an equation which can be expanded in integral powers of $z$ and
$w$. Applying the usual contour deformation argument we obtain
\begin{eqnarray}
  J(z) &=& \mathopen: \xi(z)\eta(z) \mathclose:_\lambda
  + \lambda z^{-1}, \\
  T(z) &=& \mathopen: \partial\xi(z)\eta(z) \mathclose:_\lambda
  + \frac{\lambda(\lambda-1)}{2} z^{-2}.
\end{eqnarray}
In the normal ordered products all annihilation modes
$\xi_{m-\lambda}$ and $\eta_{m-1+\lambda}$ with $m>0$ are moved to the
right.  We denote the vacuum state with respect to this normal
ordering by $\sigma_\lambda$. It has $U(1)$-charge $\lambda$ and
conformal weight
\begin{equation}
  h_\lambda = \frac{\lambda(\lambda-1)}{2}.
  \label{eq:h-twisted}
\end{equation}
Shifting the value of $\lambda$ by one does not change the boundary
conditions and results in equivalent expressions for $J$ and $T$ since
the shift in the scalar terms is compensated by a change in the normal
ordering prescription.  Thus, $\sigma_\lambda$ and
$\sigma_{\lambda+1}$ are in the same twisted sector ${\cal
  M}_\lambda$. The twist field $\sigma_\lambda$ with $0\leq\lambda<1$
is the ground state of the twisted sector. The other fields
$\sigma_{\lambda+n}$ are excited twist fields which can be obtained
{}from the ground states by acting with creation modes $\xi_{m-\lambda}$
and $\eta_{m-1+\lambda}$ with $m\leq0$.  When $\lambda\not\in\Bbb Z$
there are no zero modes of $\xi$ and $\eta$ and the ground state is
non-degenerate, while for $\lambda\in\Bbb Z$ we obtain the vacuum
representation with ground states $\Omega=\sigma_0$ and
$\xi_0\Omega=\sigma_1$.  To count the operator content of the various
sectors we introduce the $U(1)\times{\rm Vir}$ character
\begin{eqnarray}\label{eq:xieta-char}
  d_{\mu,\lambda}(\tau) &=&
  \tr_{{\cal M}_\lambda}\left(
  \e^{2\pi i\mu J_0} q^{L_0-\frac{c}{24}} \right) \nonumber\\*
  &=& \e^{-2\pi i\mu\lambda} q^{(1-6\lambda(1-\lambda))/12}
  \prod_{n=1}^\infty \left( 1 + \e^{2\pi i\mu} q^{n+\lambda-1} \right)
  \left( 1 + \e^{-2\pi i\mu} q^{n-\lambda} \right) \\
  &=& \eta(\tau)^{-1} \e^{-2\pi i\mu\lambda}
  \sum_{m\in{\Bbb Z}} \e^{2\pi i\mu m}
  q^{\frac12(m+\lambda-\frac12)^2}, \nonumber
\end{eqnarray}
where $q=\exp(2\pi i\tau)$ and $\eta(\tau)$ is the Dedekind function,
\begin{equation}
  \eta(\tau) = q^{\frac{1}{24}} \prod_{n=1}^\infty (1-q^n).
\end{equation}
For the full theory we have to take into account both the left and
right movers. We shall exclusively consider diagonal theories so that
the contribution to the partition function arising from a sector with
boundary conditions $\mu$ and $\lambda$ is
\begin{equation}
  D_{\mu,\lambda}(\tau) = |d_{\mu,\lambda}(\tau)|^2.
\end{equation}
These $D$ functions also arise as
determinants of the Laplacian in the calculation of a free field
partition function with boundary
conditions $-\exp(2\pi i\mu)$ and $-\exp(2\pi i\lambda)$, see
e.g. \cite{IDro1}.

The partition function of the $(\xi,\eta)$ system twisted by a cyclic
subgroup of $U(1)$
is obtained by summing over all (twisted) sectors and keeping only the
twist-invariant states.
If we twist by a cyclic group ${\cal C}_{2N}$ of {\em even\/} order, only
bosonic states survive in the partition function which can then be
obtained by summing the contributions of the left and right movers
over all boundary conditions,
\begin{equation}\label{eq:Znpartfn}
  Z_{N}(\tau) =
  \frac1{2N} \sum_{k,l=0}^{2N-1} D_{k/(2N),l/(2N)}(\tau)
  = \frac1{|\eta(\tau)|^2}
\sum_{n=0}^{4N^2-1} \left| \Theta_{n,2N^2}(\tau) \right|^2.
\end{equation}
The last equality expresses the partition function as a sum of theta
functions (see Appendix~\ref{sec:theta}),
\begin{equation}
  \Theta_{n,m}(\tau) = \sum_{k\in{\Bbb Z}} q^{m(k+\frac{n}{2m})^2}.
\end{equation}
Defining as in \cite{IDro1} the Coulomb gas partition function at
radius $\rho$ as
\begin{equation}\label{eq:CGZ}
  Z(\rho,\tau) = \frac1{|\eta(\tau)|^2} \sum_{m,n\in\Bbb{Z}}
  q^{\frac14(m\rho+n/\rho)^2}\bar q^{\frac14(m\rho-n/\rho)^2},
\end{equation}
we find
\begin{equation}
  Z_N(\tau) = Z(N\sqrt2, \tau).
\end{equation}
In the normalisation of \cite{Gins1} this corresponds to a Coulomb gas
partition function at radius $r=2N$.
The partition function is modular invariant as can be seen from the
transformation formulae for theta functions.

If we twist by a cyclic group ${\cal C}_N$ of {\em odd\/} order $N$ the
partition function
\begin{equation}
  Z^{NS}_N(\tau) = \frac1N \sum_{k,l=0}^{N-1}
  D_{\frac{k}{N}, \frac{l}{N}}(\tau)
\end{equation}
obtained by summing over ${\cal C}_N$ boundary conditions still
contains fermions. It is invariant under the subgroup of the modular
group generated by $T$ and $ST^2S$. The fermions are Neveu-Schwarz fields,
single-valued both on the plane and on the cylinder. To obtain a
modular invariant partition function we have to consider the possible spin
structures,
\begin{eqnarray}
  \tilde Z^{NS}_N(\tau) &=& \frac1N \sum_{k,l=0}^{N-1}
  D_{\frac{k}{N}+\frac12, \frac{l}{N}}(\tau), \\
  Z^{R}_N(\tau) &=& \frac1N \sum_{k,l=0}^{N-1}
  D_{\frac{k}{N}, \frac{l}{N}+\frac12}(\tau), \\
  \tilde Z^{R}_N(\tau) &=& \frac1N \sum_{k,l=0}^{N-1}
  D_{\frac{k}{N}+\frac12, \frac{l}{N}+\frac12}(\tau).
\end{eqnarray}
In the Ramond sector the fermions are double-valued on the plane
and the cylinder. The partition functions $\tilde Z^{NS}_N$ and $\tilde
Z^{R}_N$ have an additional insertion of the chirality operator
$(-1)^F$.
Summing over all spin structures results in the partition function of
the bosonic ${\cal C}_{2N}$ model,
\begin{equation}
  Z_N(\tau) = Z^{NS}_N(\tau) + \tilde Z^{NS}_N(\tau) +
  Z^R_N(\tau) + \tilde Z^R_N(\tau),
\end{equation}
expressing the fact that twisting by ${\Bbb Z}_2 = \{ 1, (-1)^F \}$
results in ${\cal C}_{2N} = {\Bbb Z}_2 \times {\cal C}_{N}$ for odd $N$.

\subsection{Bosonisation}
\label{sec:boson}

The connection between the $(\xi,\eta)$-system and a Coulomb gas goes
beyond an equality of partition functions.
It extends to correlation functions and operators as well.
To go to the Coulomb gas formalism we bosonise the chiral $(\xi,\eta)$
system using the natural $U(1)$-current (\ref{eq:U1-curr}),
\begin{equation}
  J(z) = i\partial x(z),
\end{equation}
where $x(z)$ is (the left moving component of) a free scalar field with
propagator
\begin{equation}
  \langle x(z)x(w)\rangle = -\ln(z-w).
\end{equation}
Because of the logarithmic propagator $x(z)$ is not itself a
Virasoro primary field but derivatives and Wick ordered exponentials
of $x(z)$ are.
The holomorphic ghost fields can be realised as
\begin{equation}
  \xi(z) = \mathopen:\e^{ix(z)}\mathclose:, \qquad
  \eta(z) = \mathopen:\e^{-ix(z)}\mathclose:,
\end{equation}
such that the stress tensor (\ref{eq:Txieta}) takes the familiar
Feigin-Fuchs form,
\begin{equation}
  T(z) = -\frac12 (\partial x(z))^2 + \frac{i}{2}\partial^2 x(z).
\end{equation}
The twist fields $\sigma_\lambda$ are represented by the vertex
operators $\mathopen:\exp(i\lambda x(z))\mathclose:$ with conformal
weight $\lambda(\lambda-1)/2$.
The twisted sector ${\cal M}_\lambda$ is given as the sum of Fock
spaces ${\cal F}_\mu$ with $\mu\in\lambda+\Bbb Z$.
This reproduces exactly the characters,
correlation functions and operator formalism of the
$(\xi,\eta)$ system. We proceed analogously to define anti-chiral
fields $\bar x, \bar\xi, \bar\eta$.

In the Coulomb gas picture one considers the free scalar field
$X(z,\bar z) = ( x(z) + \bar x(\bar z) )/\sqrt2$
with propagator $\langle X(z,\bar z)X(w,\bar w)\rangle = -\ln|z-w|$.
It is assumed to be an angular variable, $X \equiv X+2\pi\rho$, in
other words, it is compactified on a circle of radius $\rho$.
The basic fields are the electro-magnetic operators given by vertex operators
$V_{\lambda,\bar\lambda} = \mathopen:\exp(i\lambda
x(z)+i\bar\lambda\bar x(\bar z))\mathclose:$ such that they are
invariant under a shift $X \to X+2\pi\rho$ and that $X$ has a
discontinuity of $2\pi\rho m$ around a vertex operator
$V_{\lambda,\bar\lambda}$. This implies that the left and right
$U(1)$-charges are quantised as
\begin{displaymath}
  \lambda = \frac{1}{\sqrt2}\left( \frac{n}{\rho}+m\rho \right),
  \qquad
  \bar\lambda = \frac{1}{\sqrt2}\left( \frac{n}{\rho}-m\rho \right),
\end{displaymath}
with integers $m$ and $n$. All other fields are obtained by
multiplying with derivatives of the scalar field.  The states are
given by Fock spaces ${\cal F}_{\lambda,\bar\lambda}$ with
$V_{\lambda,\bar\lambda}$ corresponding to the momentum ground states.

Setting now $\rho=N\sqrt2$ for the ${\cal C}_{2N}$ model we obtain
$4N^2$ sectors ${\cal H}_j$ each being a sum of Fock spaces ${\cal
  F}_{\lambda,\bar\lambda}$ with left and right $U(1)$-charges
\begin{equation}\label{eq:xieta-charges}
  \lambda = 2Nn + \frac{j}{2N}, \qquad
  \bar\lambda = 2N\bar n + \frac{j}{2N}, \qquad n,\bar n\in{\Bbb Z}.
\end{equation}
Each sector ${\cal H}_j$ contributes a term $|\eta(\tau)^{-1}
\Theta_{j-N^2,2N^2}(\tau)|^2$ to the partition function
(\ref{eq:Znpartfn}).  The set of charges (\ref{eq:xieta-charges}) is the same
as for the ${\cal C}_{2N}$ twisted $(\xi,\eta)$ system.  We thus have
a one-to-one correspondence of the ${\cal C}_{2N}$ model and the
Coulomb gas model at radius $\rho$ and central charge $c=-2$.
This correspondence is not just at the level of correlators and the
partition function but extends to the operator formalism as well.

Due to the conserved $U(1)$-currents $J, \bar J$ the $U(1)$-charges
behave additively under operator products,
\begin{displaymath}
  V_{\lambda,\bar\lambda}(z,\bar z) {\cal F}_{\mu,\bar\mu}
  \subset {\cal F}_{\lambda+\mu,\bar\lambda+\bar\mu}.
\end{displaymath}
Hence, the fusion rules respect the charges, ${\cal
  H}_j\times{\cal H}_k = {\cal H}_{j+k}$.  But because of the
Feigin-Fuchs form of the stress tensor the $U(1)$-current is not a
Virasoro primary field,
\begin{equation}
  T(z) J(w) = \frac{-1}{(z-w)^3} + \frac{J(w)}{(z-w)^2} +
  \frac{\partial J(w)}{z-w} + O(1).
\end{equation}
As a consequence the system has a charge asymmetry, $J_0^\dagger = 1 -
J_0$.  This can be seen by taking the adjoint of the $[T,J]$
commutator \cite{FMSh1} or by looking for a definition of adjoint for
$J$ leaving the Feigin-Fuchs form of the stress-tensor invariant
\cite{Feld1}.  The charge asymmetry implies anomalous charge
conservation in correlation functions,
\begin{equation}\label{eq:chcon}
  \left\langle V_{\lambda_1,\bar\lambda_1}(z_1,\bar z_1) \cdots
    V_{\lambda_n,\bar\lambda_n}(z_n,\bar z_n) \right\rangle = 0,
    \qquad\hbox{unless $\Sigma \lambda_i = \Sigma \bar\lambda_i = 1$}.
\end{equation}
Translated to the fermionic formulation this means that correlators
are non-vanishing only if they contain exactly one unpaired $\xi$ and
$\bar\xi$ field. In particular, we have two degenerate vacua, the
M\"obius invariant vacuum $\Omega$ and $\xi_0\bar\xi_0\Omega$ such
that $\langle\Omega|\xi_0\bar\xi_0|\Omega\rangle = 1$.  As a further
consequence of the charge asymmetry it is not possible to have an
inner product $(\cdot,\cdot)$ on the vacuum sector compatible with
$L_n^\dagger=L_{-n}$. In fact, $(v,w)$ can be non-vanishing only if
$v\in{\cal H}_j$ and $w\in{\cal H}_{2N-j}$.  A further problem is the
appearance of fields which are neither primary fields nor descendents
of a primary field. The simplest example is the current $J$ which gets
mapped into the conformal family of the identity by the action of $T$.
The representation of the Virasoro algebra generated from the current
$J$ is thus a reducible but indecomposable representation.

The lack of an inner product and the appearance of reducible
representations of the Virasoro algebra is a result of the charge-asymmetry
and does not occur in Virasoro minimal models.  In the
Coulomb gas construction of Virasoro minimal models one therefore
introduces screening charges which change the $U(1)$-charge but
commute with the Virasoro algebra. The physical states of the minimal
model are defined through a BRST resolution \cite{Feld1}. This
identifies the two ground states and removes the reducible
representations of the Virasoro algebra leaving only the field content
of the minimal model. In our case the screening charge corresponds to
$\eta_0$ and the usual ``physical'' space $\mathop{\rm
  ker}\eta_0/\mathop{\rm im}\eta_0$ is trivial, as observed in
\cite{Sale1}.  We can, however, take the kernel of $\eta_0$ without
going to the quotient. This defines a ``small'' algebra which we
discuss in the next section.

\section{The ``small'' algebra}
\label{sec:proj}

It was noted in \cite{FMSh1} that the $(\xi,\eta)$ system contains a
``small'' algebra generated by $\partial\xi$ and $\eta$. In this
section we show that this ``small'' algebra can be characterised as
the kernel of $\eta_0$ on the space of states for the $(\xi,\eta)$
system. It has a unique ground state and is completely reducible into
Virasoro highest highest weight representations.

Let us return to the chiral untwisted $(\xi,\eta)$ system.  Its space
of states is spanned by lexicographically ordered monomials of the
creation modes $\xi_m, m\geq0$ and $\eta_m, m>0$ acting on the
M\"obius invariant vacuum $\Omega$. The operator
$\eta_0$ vanishes on all states not containing the mode $\xi_0$ since
$\eta_0$ and $\xi_0$ are conjugate operators.
The correspondence of states and fields is
\begin{displaymath}
  \xi_{-m_1}\cdots\xi_{-m_r} \eta_{-n_1-1}\cdots\eta_{-n_s-1} \Omega
  \longleftrightarrow
  \mathopen: \frac{\partial^{m_1}\xi(z)}{m_1!}\cdots
  \frac{\partial^{m_r}\xi(z)}{m_r!}
  \frac{\partial^{n_1}\eta(z)}{n_1!}\cdots
  \frac{\partial^{n_s}\eta(z)}{n_s!}\mathclose:.
\end{displaymath}
The zero mode $\xi_0$ can only be generated by the field $\xi(z)$ and
is not present in derivatives of $\xi(z)$. Hence, if we
start off with states not containing $\xi_0$ the zero mode does not
get generated. We can therefore consistently restrict to the kernel of
$\eta_0$. This space is spanned by ordered monomials of the negative modes
$\xi_m, \eta_m, m<0$ acting on the vacuum $\Omega$.
All the fields can be generated by taking derivatives and normal
ordered products of $\partial\xi$ and $\eta$.
This is the ``small'' algebra of \cite{FMSh1}.  For any state $v$ in
the ``small'' algebra there is a second state $\xi_0 v$ in the
$(\xi,\eta)$ system.

Both $\partial\xi$ and $\eta$ are Virasoro primary fields
of dimension one and have $U(1)$-charges $\pm1$.  We can put
them on equal footing by writing them as the two components of a
fermionic dimension one field $\psi$,
\begin{equation}
  \psi^+(z) = \eta(z), \qquad \psi^-(z) = \partial\xi(z).
\end{equation}
The anti-commutator (\ref{eq:xieta-acr}) reads then
\begin{equation}\label{eq:psi-acr}
  \{ \psi^\alpha_m, \psi^\beta_n \} = m J^{\alpha\beta} \delta_{m+n},
\end{equation}
where the anti-symmetric tensor $J$ is defined as $J^{+-} = -J^{-+} =
1$.  For the stress tensor we find
\begin{equation}\label{eq:Tpsi}
  T(z) = \frac12 J_{\alpha\beta}
  \mathopen:\psi^\alpha(z)\psi^\beta(z)\mathclose: ,
\end{equation}
where $J_{\alpha\beta}$ is the inverse of $J^{\alpha\beta}$.
The space of states of the ``small'' algebra is given by ordered
monomials of negative modes of $\psi^+$ and $\psi^-$ acting on the
vacuum. Because of Fermi statistics each mode appears at most once.

Correlation functions of the $\psi$ field can easily be calculated
using a fermionic version of Wick's theorem and are given
by the Pfaffian
\begin{equation}
  \langle\psi^{\alpha_1}(z_1)\cdots\psi^{\alpha_{2n}}(z_{2n})\rangle =
  \Pf\left( \frac{J^{\alpha_i\alpha_j}}{(z_i-z_j)^2} \right) .
\end{equation}

The ``small'' algebra is determined completely by the anti-commutators
(\ref{eq:psi-acr}) of the generating fermion fields and the expression
(\ref{eq:Tpsi}) for the stress tensor.  It has a global $SL(2)$
symmetry, which we will call isospin, acting on the basic fermion
field $\psi$ according to
\begin{displaymath}
  g\colon\pmatrix{\psi^+ \cr \psi^- } \mapsto
  \pmatrix{g_{11} & g_{12} \cr g_{21} & g_{22}}
  \pmatrix{\psi^+ \cr \psi^- }
\end{displaymath}
with the matrix $g_{ij}$ in $SL(2)$.
We can choose an explicit basis of the Lie algebra $sl(2)$, acting on
the basic fermion field $\psi$ as
\begin{equation}
    J^0 \psi^\pm = \pm\frac12 \psi^\pm, \qquad
    J^\pm \psi^\pm = 0, \qquad
    J^\pm \psi^\mp = \psi^\pm,
\end{equation}
and on their products in the usual way.  The basic fermion field
$\psi$ thus transforms as the isospin $\frac12$ representation of
$sl(2)$. The original $U(1)$-symmetry of the $(\xi,\eta)$ system is
the $U(1)$-subgroup generated by $J^0$.

The space of states will decompose into representations of
$SL(2)\times{\rm Vir}$. The Virasoro primary states in the
isospin $j$ multiplet can be obtained explicitly as
\begin{equation}\label{eq:phijm}
  \phi^{j,m} =
  \psi_{-2j}^{(+} \cdots \psi_{-j+m}^{+}
  \,\psi_{-j+m+1}^{-} \cdots \psi_{-1}^{-)} \Omega,
  \label{phijm}
\end{equation}
where $j\in\frac12\Bbb Z, m=-j,-j+1,\ldots,j$ and we symmetrise over
the signs. The generating fermions $\psi^\pm$ form the isospin
$\frac12$ multiplet. The structure constants for the
$\phi^{j,m}$ fields are invariant under $SL(2)$, that is, they are
Clebsch-Gordan coefficients up to some normalisation factor.

The fields with half-integral isospin are fermionic while the fields
with integral isospin are bosonic.
The bosonic sector is generated by the isospin 1 fields,
\begin{eqnarray}
  W^+ = \phi^{1,1} &=& \partial\psi^+ \psi^+, \\
  W^0 = \phi^{1,0} &=& \frac12 \left( \partial\psi^+ \psi^- +
  \partial\psi^- \psi^+ \right), \\
  W^- = \phi^{1,-1} &=& \partial\psi^- \psi^-.
\end{eqnarray}
As they are of dimension three they do not form a Kac-Moody algebra
but a W-algebra \cite{Kaus1} with operator product expansion
\begin{eqnarray}
  T(z) T(w) &\sim&\frac{-1}{(z-w)^4} + \frac{2T(w)}{(z-w)^2} +
  \frac{\partial T(w)}{z-w}, \\
  T(z) W^j(w) &\sim& \frac{3W^j(w)}{(z-w)^2} +
  \frac{\partial W^j(w)}{z-w}, \\
  W^i(z) W^j(w) &\sim& g^{ij} \Biggl( \frac{1}{(z-w)^6}
  - 3\frac{T(w)}{(z-w)^4} +
  -\frac32 \frac{\partial T(w)}{(z-w)^3} +
  \nonumber\\*&&\qquad
  +\frac32\frac{\partial^2 T(w)}{(z-w)^2}
  - 4 \frac{(T^2)(w)}{(z-w)^2}
  +\frac16 \frac{\partial^3 T(w)}{z-w}
  -4 \frac{\partial(T^2)(w)}{z-w} \Biggr)
  \\*&&
  -5f^{ij}_k \Biggl( \frac{W^k(w)}{(z-w)^3} +
  \frac12 \frac{\partial W^k(w)}{(z-w)^2} +
  \frac{1}{25}\frac{\partial^2 W^k(w)}{z-w} +
  \frac{1}{25}\frac{(TW^k)(w)}{z-w} \Biggr)
  \nonumber,
\end{eqnarray}
where $g^{ij}$ is the metric on the isospin one representation,
$g^{+-} = g^{-+} = 2, g^{00} = -1$, and $f^{ij}_k$ are the structure
constants of $sl(2)$, such that $f^{ijk} = f^{ij}_l g^{lk}$ is totally
antisymmetric and normalised to $f^{+-0}=2$.

Each isospin $j$ multiplet appears exactly once in the ``small'' algebra
as can be seen by introducing the $sl(2)\times{\rm Vir}$
character
\begin{equation}
  \chi(\tau,z) = \tr \left( w^{J^0} q^{L_0-\frac{c}{24}} \right),
\end{equation}
where $w = \exp(2\pi iz)$.
The fields $\phi^{j,m}$ have conformal weight
$j(2j+1)$ which corresponds to $h_{2j+1,1}$ in the Kac-table for
$c=-2$. The Virasoro character for each of the $\phi^{j,m}$ is thus
\begin{displaymath}
  \chi^{\rm Vir}_{2j+1,1}(\tau) = q^{1/8} \eta(\tau)^{-1}
  \left( q^{j(2j+1)} - q^{(j+1)(2j+1)} \right)
\end{displaymath}
The sum over the isospin $j$ multiplets is then
\begin{eqnarray}\label{eq:psi-char}
  \chi(\tau,z) &=& \sum_{j\in\frac12\Bbb{Z}_{\geq0}} \sum_{k=-j}^{j}
  w^{k} \chi^{\rm Vir}_{2j+1,1}(\tau) \nonumber\\*
  &=& \eta(\tau)^{-1}
  \sum_{l=-\infty}^\infty \frac{w^{l/2} - w^{-l/2}}{w^{1/2}-w^{-1/2}}
  \, q^{(2l-1)^2/8}.
\end{eqnarray}
This agrees with the direct calculation
\begin{equation}\label{eq:GVir-char}
  \chi(\tau,w) = q^{\frac{1}{12}} \prod_{n=1}^\infty
  (1+w^{\frac12}q^n) (1+w^{-\frac12}q^n).
\end{equation}
The Ramond sector, with ground state energy $h=-1/8$,
decomposes in the same way into isospin multiplets with each multiplet
appearing exactly once. The Virasoro primary states $\chi^{j,m}$ in the
isospin $j$ multiplet of the Ramond sector can be obtained explicitly as
\begin{equation}\label{eq:chijm}
  \chi^{j,m} =
  \psi_{-2j+1/2}^{(+} \cdots \psi_{-j+m+1/2}^{+}
  \,\psi_{-j+m+3/2}^{-} \cdots \psi_{-1/2}^{-)} \sigma_{1/2},
  \label{chijm}
\end{equation}
where $j\in\frac12\Bbb Z, m=-j,-j+1,\ldots,j$ and we symmetrise over
the signs.
Due to the half-integral moding, $\chi^{j,m}$ has
conformal weight $2j^2-1/8$, which corresponds to $h_{2j+1,2}$ in the
Kac-table. Their Virasoro characters are
\begin{displaymath}
  \chi^{\rm Vir}_{2j+1,2}(\tau) = \eta(\tau)^{-1}
  \left( q^{2j^2} - q^{2(j+1)^2} \right)
\end{displaymath}
leading to the $sl(2)\times{\rm Vir}$ character for the Ramond sector
\begin{eqnarray}
  \chi^{R}(\tau,z) &=& \sum_{j\in\frac12\Bbb{Z}_{\geq0}} \sum_{k=-j}^{j}
  w^{k} \chi^{\rm Vir}_{2j+1,2}(\tau) \nonumber\\*
  &=& \eta(\tau)^{-1}
  \sum_{l=-\infty}^\infty w^{l/2} q^{l^2/2} \\*
  &=& q^{-\frac{1}{24}} \prod_{n=1}^\infty
  (1+w^{\frac12}q^{n+\frac12}) (1+w^{-\frac12}q^{n+\frac12}).
  \nonumber
\end{eqnarray}
Starting with the ``small'' algebra one can repeat the orbifold
constructions of the previous section. The characters of the twisted
sectors will be unchanged. However, the character (\ref{eq:psi-char})
of the untwisted sector has no nice modular transformation
properties.
Take for example the ${\cal C}_2$ orbifold. The characters
of the four (chiral) sectors are
\begin{equation}
\begin{array}{rcl@{\qquad}rcl}
  \chi_0(\tau) &=& \displaystyle
  \frac12 \left( \frac{\Theta_{1,2}(\tau)}{\eta(\tau)} +
  \eta(\tau)^2 \right), &
  \chi_1(\tau) &=& \displaystyle
  \eta(\tau)^{-1} \Theta_{0,2}(\tau), \\
  \chi_2(\tau) &=& \displaystyle
  \frac12 \left( \frac{\Theta_{1,2}(\tau)}{\eta(\tau)} -
  \eta(\tau)^2 \right), &
  \chi_3(\tau) &=& \displaystyle
  \eta(\tau)^{-1} \Theta_{2,2}(\tau).
\end{array}
\end{equation}
The partition function should be the diagonal combination of these
characters. But the characters $\chi_0$ and $\chi_2$ are sums of a
modular function and a modular form of weight one. Defining formally an
action of the modular group on these characters results in characters
having an additional prefactor of $\log q = 2\pi i\tau$,
\begin{eqnarray}
  \chi_0(-1/\tau) &=& \displaystyle
  \frac14 \chi_1(\tau) - \frac14 \chi_3(\tau)
  -\frac{i\tau}{2}\eta(\tau)^2, \\
  \chi_1(-1/\tau) &=& \displaystyle
  \chi_0(\tau) + \frac12\chi_1(\tau) + \chi_2(\tau)
  + \frac12\chi_3(\tau), \\
  \chi_2(-1/\tau) &=& \displaystyle
  \frac14 \chi_1(\tau) - \frac14 \chi_3(\tau)
  +\frac{i\tau}{2}\eta(\tau)^2, \\
  \chi_3(-1/\tau) &=& \displaystyle
  -\chi_0(\tau) + \frac12\chi_1(\tau) - \chi_2(\tau)
  + \frac12\chi_3(\tau).
\end{eqnarray}
The situation for a general ${\cal C}_{2N}$ orbifold is analogous. The
twisted sectors are the same as for the $(\xi,\eta)$-system while the
untwisted sector splits into linear combinations of $\eta(\tau)^2$ and
terms involving theta functions,
$\eta(\tau)^{-1}\Theta_{(2k-N)N,2N^2}(\tau)$ for $k=0,\ldots,2N-1$.
For a discussion of these issues and an attempt at a physical
interpretation and construction of modular invariant partition
functions see \cite{Floh2}.

\subsection{Bosonisation and fusion}
\label{sec:mini}

The ${\cal C}_2$-twisted $(\xi,\eta)$ system and ``small'' algebra
share many features with minimal models: The central charge $c=-2$ can
be obtained from the minimal model formula
\begin{equation}
  c = 1 - 6\frac{(p-p')^2}{pp'}
\end{equation}
for $(p,p') = (1,2)$.
The conformal weights can all be found in the Kac-table as
\begin{equation}
  h_{n,n'} = \frac{(2n-n')^2-1}{8}
\end{equation}
corresponding to the $U(1)$-charges $\lambda_{n,n'} = -n + (n'+1)/2$.
The ${\cal C}_{2N}$ model for higher twists, $N>1$, can not be
interpreted in minimal model language since they involve fields with
non-integral labels in the Kac-table.  However, neither ${\cal C}_2$
model should strictly be called a $(1,2)$ minimal model for the
following reasons. Firstly, both models contain an infinite number of
Virasoro primary fields in contrast to the usual minimal models.
Secondly, the treatment of minimal models following Belavin, Polyakov and
Zamolodchikov \cite{BPZ} supposes a specific embedding structure of
Virasoro modules and Fock spaces which is only the case for $p,p'>1$
and relatively prime \cite{FeFu2}. And thirdly, both models contain
fields which belong to reducible but indecomposable representations of
the Virasoro algebra.

In the $(\xi,\eta)$ system the operator product expansions
of the Ramond-fields can easily be calculated in the bosonised
formalism and one obtains, for example,
\begin{eqnarray}
  \chi^{0,0}(z) \chi^{0,0}(w) &\sim& (z-w)^{\frac14}
  \Biggl\{ \xi(w) +
    \frac12 \psi^-(w) (z-w) +
    \cdots\Biggr\}
  \\
  \chi^{0,0}(z) \chi^{\frac12,\frac12}(w) &\sim& (z-w)^{-\frac14}
  \Biggl\{ \id +
    \frac12 J(w) (z-w)
    \nonumber\\*&&+
    \left(\frac14 T(w) +
      \frac{1}{8} \partial J(w) \right) (z-w)^2
    \\*&&+
    \left(\frac{5}{48} \partial T(w) +
      \frac{1}{32} \partial^2J(w) +
      \frac{1}{24} W^0(w) \right) (z-w)^3 +
    \cdots\Biggr\}
  \nonumber
\end{eqnarray}
On the right hand side we find the conformal families generated by the
Virasoro primary fields of
the identity, $\phi^{0,0}=\id$, the isospin $1/2$ field
$\phi^{1/2,-1/2}=\psi^-=\partial\xi$ and the isospin 1 field
$\phi^{1,0}=W^0$. But we also produce the fields $\xi$ and $J$ generating
reducible representations, such that
\begin{equation}
  L_1 J = -\Omega, \qquad \left(L_{-2}-\frac12 L_{-1}^2\right) J = W^0,
  \qquad L_{-1} \xi = \psi^-
\end{equation}
on the states. The fact that the identity and $J$ are coupled in one
reducible representation makes it possible for the identity to appear in
the operator product of two primary fields of different conformal
weight.

Correlation functions for the ``small'' algebra can be calculated from
null vector constraints:
The states $\chi^{j,m}$ have
null vectors at level $2(2j+1)$ and as a consequence the 4pt function
$\langle\chi^{j_1,m_1}(z_1)\cdots\chi^{j_4,m_4}(z_4)\rangle$ satisfies
a Fuchsian differential equation. For the full, non-chiral 4pt
function we have to take linear combinations of the chiral and
anti-chiral solutions to the differential equations such that the full
correlator has no monodromy.  The generic solution of the differential
equation and monodromy conditions is given by the integral
representation of Dotsenko and Fateev \cite{DFat1}.  Consider now
specifically the 4pt function for the ground state of the Ramond
sector, $\mu(z,\bar z) = \chi^{0,0}(z)\bar\chi^{0,0}(\bar z)$.  The
4pt correlator $\langle\mu\mu\mu\mu\rangle$ calculated according to
\cite{DFat1} vanishes identically.  This can be traced back to the
Fuchsian differential equation satisfied by the correlators acquiring
exponents differing by an integer. Hence some solutions of the
differential equation become degenerate.  The linear combination of
these degenerate solutions appearing in the correlation function is
such that they cancel identically. The generic solution of
\cite{DFat1} is only valid in the non-degenerate case.  To get a
non-zero result one has to consider the $c\rightarrow-2$ limit and
obtains \cite{Sale1}
\begin{equation}\label{eq:4mucorr}
  \langle\mu\mu\mu\mu\rangle =
  |z_{12}z_{34}|^{\frac12} |x(1-x)|^{\frac12}
  \left( F(x)\overline{F(1-x)} + F(1-x)\overline{F(x)} \right),
\end{equation}
where $x = (z_{12}z_{34})/(z_{13}z_{24})$ is the cross-ratio and
$F(x)$ is the hypergeometric function
\begin{displaymath}\textstyle
  F(x) = {}_2F_1(\frac12,\frac12;1;x).
\end{displaymath}
This result can be obtained without taking limits in the ${\cal C}_2$
twisted ``small'' algebra using the techniques of \cite{DFMS1} or by
directly solving the differential equation and monodromy conditions at
$c=-2$ \cite{Gura1}.

Conformal invariance fixes 4pt functions up to a function of the
cross-ratios $x, \bar x$. This function can be written as a sum over
products of a chiral conformal block
times an anti-chiral conformal block corresponding to the various
fields propagating in the intermediate channel. If we keep
$z_1$ near $z_2$ and $z_3$ near $z_4$ and pull the two pairs apart,
corresponding to the limit $x\to0$,
the only fields appearing in the intermediate channel of a conformal
block for $\langle\mu(z_1)\cdots\mu(z_4)\rangle$
are those which
appear in the fusion of $\chi^{0,0}$ with itself.
The 4pt function (\ref{eq:4mucorr}) has
logarithmic singularities at $x=0$ and $x=1$.
Analytic continuation of the hypergeometric function yields
\begin{displaymath}
  F(1-x) = -\frac{1}{\pi} \Big( \ln(x/16) F(x) + M(x) \Big),
\end{displaymath}
where $M(x)$ is some regular function vanishing at the origin.
We thus have
\begin{equation}\label{eq:4mucorr0}
  \langle\mu\mu\mu\mu\rangle =
  |z_{12}z_{34}|^{\frac12} |x(1-x)|^{\frac12}
  \left\{ C_1 |F(x)|^2 + C_2 \Big(
    \big(\ln(x)F(x)+M(x)\big)\overline{F(x)} + c.c. \Big)
  \right\},
\end{equation}
with constants $C_1, C_2$.
Conformal blocks can be calculated perturbatively with the techniques
of Appendix B of \cite{BPZ}. The first summand in (\ref{eq:4mucorr0})
can indeed be identified as the conformal block for coupling through
the vacuum.
However, the second summand has a
contribution with a $\ln(x)$ behaviour for small values of the
cross-ratio $x$.  It was argued in \cite{Gura1} that this implies the
appearance of ``logarithmic'' operators in the operator
product expansion $\chi^{0,0}(z)\chi^{0,0}(0)$. These operators
correspond to 2-dimensional Jordan cells for $L_0$ and $\bar L_0$.
The perturbative calculation of conformal blocks can be adapted to
such ``logarithmic'' operators and reproduces exactly the second
summand in (\ref{eq:4mucorr0}) if one couples through a field
$\omega(z,\bar z)$ with operator product expansions,
\begin{eqnarray}\label{eq:Tw-ope}
  T(z) \omega(w, \bar w) &\sim& \frac{1}{(z-w)^2} +
  \frac{\partial\omega(w, \bar w)}{z-w}, \\
  \label{eq:Tbw-ope}
  \bar T(\bar z) \omega(w, \bar w) &\sim& \frac{1}{(\bar z-\bar w)^2} +
  \frac{\bar\partial\omega(w, \bar w)}{\bar z-\bar w}.
\end{eqnarray}
Such a field does not exist in the twisted $(\xi,\eta)$ system nor the
twisted ``small'' algebra since there $L_0$ is always diagonalisable.
While the twisted $(\xi,\eta)$ system has well-defined fusion albeit
involving reducible representations of the Virasoro algebra, the
twisted ``small'' algebra is not closed under fusion. To define a
consistent conformal field theory one has to extend the space of
states to include two-dimensional Jordan cells for $L_0$ and $\bar
L_0$.  Let us therefore consider a representation of the chiral ``small''
algebra containing it.  We assume $L_0$ has a lowest eigenvalue on the
representation such that the energy is bounded from below.  On the
lowest energy subspace all positive modes of $\psi^\pm$ vanish and we
have $L_0 = \psi^-_0 \psi^+_0$ because of the normal ordering
prescription.  The zero modes of $\psi^\pm$ form a two-dimensional
Grassmann algebra. The lowest energy subspace will decompose into
representations of that Grassmann algebra. Its maximal indecomposable
representation is four-dimensional, spanned by
$\{\Omega,\phi^\pm,\omega\}$ such that
\begin{equation}\label{eq:zma}
  \psi_0^\mp \phi^\pm = \pm i \Omega, \qquad
  \psi_0^\pm \omega = i\phi^\pm, \qquad
  L_0 \omega = \Omega,
\end{equation}
and the other actions vanish.
The factors of $i$ are chosen to agree with the conventions of the
following section.
{}From (\ref{eq:zma}) we see that $L_0$ has zero eigenvalue on the
lowest energy subspace with $(\omega,\Omega)$ forming a
two-dimensional Jordan cell. The full representation is generated from
(\ref{eq:zma}) by the free action of the negative modes of $\psi^\pm$.
For any state built on the M\"obius invariant vacuum state $\Omega$,
corresponding to a state of the ``small'' algebra,
there are three other states in the representation generated from
(\ref{eq:zma}) built on $\phi^\pm$ and $\omega$.
The maximal extension of the chiral ``small'' algebra has thus four ground
states, two bosonic states $\{\Omega,\omega\}$ in a Jordan block and
two fermion states $\phi^\pm$ of conformal weight zero.
In a non-chiral model we can have at most 16 ground states
corresponding to the product of the left times the right maximal
extension.

\section{Symplectic fermions}
\label{sec:symp}

In this section we will present a construction for the maximal
diagonal extension of the non-chiral ``small'' algebra based on
``symplectic''
fermions.
Specifically we consider a free fermionic field $\Phi(z,\bar z)$ taking
values in a two-dimensional space with symplectic form
$J^{\alpha\beta}$. The action is
\begin{equation}\label{eq:action}
  S = \frac1{4\pi} \int\!{\d^2z} J_{\alpha\beta}
  \partial_\mu \Phi^\alpha \partial^\mu \Phi^\beta,
\end{equation}
where $J_{\alpha\beta}$ is the inverse of the symplectic form,
$J_{\alpha\beta}J^{\beta\gamma} = \delta_\alpha^\gamma$.
The propagator is
\begin{equation}\label{eq:prop}
  \langle \Phi^\alpha(z,\bar z) \Phi^\beta(z',\bar z') \rangle =
  - J^{\alpha\beta} \ln|z-z'|^2.
\end{equation}
The general solution of the equations of motion
is
\begin{equation}
  \Phi^\alpha(z,\bar z) = \phi^\alpha(z) + \bar\phi^\alpha(\bar z),
\end{equation}
where $\phi^\alpha$ and $\bar\phi^\alpha$ are arbitrary functions of
their respective arguments, subject only to periodicity or boundary
conditions. Invariance of the action under $\sigma\mapsto\sigma+2\pi$
implies the mode expansion
\begin{equation}
  \phi^\alpha(z) = \phi^\alpha_0 -i \psi^\alpha_0 \ln z +
  i \sum_{n\neq0} \frac{\psi^\alpha_n}n z^{-n},
\end{equation}
and analogously for $\bar\phi^\alpha(\bar z)$.  The chiral fields
$\phi^\alpha, \bar\phi^\alpha$ are not completely independent but are
coupled through their zero-modes. The periodicity condition
$\Phi^\alpha(\e^{2\pi i}z,\e^{-2\pi i}\bar z)=\Phi^\alpha(z,\bar z)$
implies $\psi^\alpha_0 = \bar\psi^\alpha_0$ on the physical states.
Thus, the zero-mode algebra acting on physical states involves only
the combinations
\begin{equation}
  \Phi^\alpha_0 = \phi^\alpha_0 + \bar\phi^\alpha_0, \qquad
  \Psi^\alpha_0 = \frac12( \psi^\alpha_0 + \bar\psi^\alpha_0).
\end{equation}
The mode expansion for the non-chiral field
$\Phi^\alpha$ can then be written
\begin{equation}
  \Phi^\alpha(z,\bar z) = \Phi^\alpha_0 -i \Psi^\alpha_0 \ln|z|^2 +
  i \sum_{n\neq0} \frac{\psi^\alpha_n}n z^{-n} +
  \frac{\bar\psi^\alpha_n}n \bar z^{-n}.
\end{equation}
The modes $\phi^\alpha_0, \psi^\alpha_n$ for $n<0$ are creation modes
while $\psi^\alpha_n$ for $n\geq0$ are annihilation modes. We use the
fermionic normal ordering prescription
\begin{equation}\label{eq:nord}
  \mathopen: \phi^\alpha_0 \psi^\beta_0 \mathclose: =
  -\mathopen: \psi^\beta_0 \phi^\alpha_0 \mathclose: =
  \phi^\alpha_0 \psi^\beta_0, \qquad
  \mathopen: \psi^\alpha_m \psi^\beta_n \mathclose: =
  \left\{
    \begin{array}{cl}
      \psi^\alpha_m \psi^\beta_n & \hbox{for $m<0$}, \\
      -\psi^\beta_n \psi^\alpha_m & \hbox{for $m>0$}.
    \end{array}
  \right.
\end{equation}
The propagator (\ref{eq:prop}) then implies the anti-commutators
\begin{equation}
  \{ \psi^\alpha_m, \psi^\beta_n \} = m J^{\alpha\beta} \delta_{m+n},
  \qquad
  \{ \phi^\alpha_0, \psi^\beta_0 \} =
  \{ \Phi^\alpha_0, \Psi^\beta_0 \} = i J^{\alpha\beta}.
\end{equation}
The logarithmic propagator shows that correlators involving $\Phi$
require a careful treatment of the infra-red and ultra-violet
divergences. However, derivatives of $\Phi$ are without problems and
reproduce the results of the previous section.
The action is invariant under a constant shift of the fermion fields,
$\delta\Phi^\alpha(z,\bar z) = \beta^\alpha$,
resulting in the conserved currents
\begin{equation}
  \psi^\alpha(z) = i\partial \Phi^\alpha(z,\bar z), \qquad
  \bar\psi^\alpha(\bar z) = i\bar\partial \Phi^\alpha(z,\bar z),
\end{equation}
which are the dimension one fermions of the previous section. Their
propagators are
\begin{equation}
  \begin{array}{rclrcl}
    \langle\psi^\alpha(z)\psi^\beta(w)\rangle &=&
    \frac{J^{\alpha\beta}}{(z-w)^2}, &
    \langle\Phi^\alpha(z,\bar z)\psi^\beta(w)\rangle &=&
    iJ^{\alpha\beta} \frac{1}{z-w}, \\
    \langle\bar\psi^\alpha(z)\bar\psi^\beta(w)\rangle &=&
    \frac{J^{\alpha\beta}}{(\bar z-\bar w)^2}, &
    \langle\Phi^\alpha(z,\bar z)\bar\psi^\beta(\bar w)\rangle &=&
    iJ^{\alpha\beta} \frac{1}{\bar z-\bar w}, \\
    \langle\psi^\alpha(z)\bar\psi^\beta(\bar w)\rangle &=& 0,
  \end{array}
\end{equation}
showing that the ``small'' algebra is contained in the symplectic
fermion model.
The stress tensor obtained from the action is
\begin{equation}
  T_{\mu\nu} = -\frac12 J_{\alpha\beta}
  \partial_\mu\Phi^\alpha \partial_\nu\Phi^\beta +
  \frac14 g_{\mu\nu} J_{\alpha\beta}
  \partial_\lambda\Phi^\alpha \partial^\lambda\Phi^\beta.
\end{equation}
Its components in the complex basis are
\begin{equation}
  T(z) = \frac12 J_{\alpha\beta}
  \mathopen:\psi^\alpha(z) \psi^\beta(z)\mathclose:, \qquad
  \bar T(\bar z) = \frac12 J_{\alpha\beta}
  \mathopen:\bar\psi^\alpha(\bar z) \bar\psi^\beta(\bar z)\mathclose:
\end{equation}
and the central charge is $c=-2$.

The vacuum representation is generated from the vacuum $\Omega$ by the
action of the creation modes $\Phi^\alpha_0$ and $\psi^\alpha_{-n},
\bar\psi^\alpha_{-n}$ for $n>0$. The partition function is thus
$D_{0,0}(\tau)$ as for the $(\xi,\eta)$-system.  In particular, there
are four states of dimension zero, the two fermionic states
$\Phi^\alpha_0\Omega$, the vacuum state $\Omega$ and another bosonic
state
\begin{equation}
  \omega = \frac12 J_{\alpha\beta} \Phi^\alpha_0 \Phi^\beta_0 \Omega.
\end{equation}
The fermionic states $\Phi^\alpha_0\Omega$ are eigenstates of $L_0$
and $\bar L_0$ while the vacuum and $\omega$ form a
two-dimensional Jordan cell to the eigenvalue zero for $L_0$,
\begin{displaymath}
  L_0 \omega = \bar L_0 \omega = \Omega
\end{displaymath}
{}From this one finds the operator product expansion of the field
$\omega(z,\bar z)$ with the stress tensor
as in (\ref{eq:Tw-ope}) and (\ref{eq:Tbw-ope})
indicating that $\omega(z,\bar z)$ does indeed provide the
``logarithmic'' operator required in the 4pt functions of twist fields
\cite{Gura1}.

The symplectic fermions provide a non-chiral version of the maximal
extension of the ``small'' algebra discussed in the previous section.
It is non-chiral since the left and right zero modes are coupled
through the periodicity condition.  The $(\xi,\eta)$ system and the
symplectic fermions both have four ground states and
are two different subtheories of the product
theory generated by the ``chiral'' fields $\phi^\alpha(z)$ and the
``anti-chiral'' fields $\bar\phi^\alpha(\bar z)$. This product theory
is rather problematic, however, since the ``chiral'' fields are not
meromorphic functions of the coordinates $z$ but have a $\log(z)$
dependence as well.  Specifically, setting $\psi^-_0\equiv0$ and
$\bar\psi^-_0\equiv0$ removes their partners $\phi^+_0, \bar\phi^+_0$ from
the theory and
we obtain the $(\xi,\eta)$ system with the identification
$\xi = \phi^-, \eta=i\partial\phi^+$ and
$\bar\xi = \bar\phi^-, \bar\eta=i\bar\partial\bar\phi^+$.
The field $\xi$ is then a proper chiral field since the troublesome
$\log(z)$ term has been removed by the $\psi^-_0\equiv0$ condition.
If instead we take the diagonal choice,
$\psi^\pm_0\equiv\bar\psi^\pm_0$, we obtain the symplectic fermions.
Setting both $\psi^\pm_0\equiv0$ and $\bar\psi^\pm_0\equiv0$ yields
the ``small'' algebra.

\subsection{$SL(2)$ symmetry}
\label{sec:sl2}

The action (\ref{eq:action}) is invariant under $SL(2)$
transformations on the field $\Phi$. We write an infinitesimal
transformation as
\begin{equation}
  \delta\Phi^\alpha(z,\bar z) =
  -i\Lambda_a(z,\bar z) d^{a\alpha}_\beta \Phi^\beta(z,\bar z),
\end{equation}
where the $d^{a\alpha}_\beta$ are representation matrices for the
two-dimensional representation of $sl(2)$, see Appendix \ref{sec:conv}
for conventions used. Noether's theorem then leads to currents
\begin{equation}
  J^a(z,\bar z) = \frac{i}{2} d^a_{\alpha\beta}
  \mathopen:\Phi^\alpha(z,\bar z) \psi^\beta(z)\mathclose:, \qquad
  \bar J^a(z,\bar z) = \frac{i}{2} d^a_{\alpha\beta}
  \mathopen:\Phi^\alpha(z,\bar z) \bar\psi^\beta(\bar z)\mathclose:,
\end{equation}
where we wrote explicitely the dependence on the holomorphic and
anti-holomorphic coordinate and
$d^a_{\alpha\beta} = d_\alpha^{a\gamma} J_{\gamma\beta}$.
The currents are conserved,
\begin{equation}
  \bar\partial J^a + \partial \bar J^a = 0.
\end{equation}
Since the currents contain the field $\Phi$ directly they require
careful renormalisation beyond the normal ordering prescription
(\ref{eq:nord}). However the charges are well-defined,
\begin{eqnarray}
  Q^a &=& \frac{1}{2\pi i} \int (J^a \d{}z - \bar J^a \d{}\bar z)
  \nonumber\\*
  &=& d^a_{\alpha\beta} \left\{ i \Phi^\alpha_0 \Psi^\beta_0 +
  \sum_{n=1}^\infty \left( \frac{\psi^\alpha_{-n}\psi^\beta_n}{n} +
  \frac{\bar\psi^\alpha_{-n}\bar\psi^\beta_n}{n} \right) \right\}.
  \nonumber
\end{eqnarray}
and satisfy an $sl(2)$ algebra,
\begin{equation}\label{eq:chargealg}
  {}[ Q^a, Q^b ] = f^{ab}_c Q^c.
\end{equation}
One can check explicitly that the fermion field $\Phi$ transforms under
the two-dimensional representation of $SL(2)$,
\begin{equation}
  [Q^a, \Phi^\alpha(z,\bar z) ] = d^{a\alpha}_\beta
  \Phi^\beta(z,\bar z),
\end{equation}
and that the fundamental propagator (\ref{eq:prop}) is invariant under
the $SL(2)$ transformations generated by $Q^a$. Thus, the $SL(2)$
symmetry of symplectic fermion model (\ref{eq:action}) remains valid
in the quantum theory.

\subsection{Orbifolds}
\label{sec:orbs}

Given a modular invariant conformal field theory with a (finite)
symmetry group $\Gamma$ one can construct another such theory as the
orbifold by $\Gamma$ \cite{DHVW1,DHVW2}.
In the Hamiltonian picture the orbifold theory is obtained by adding
twisted sectors, corresponding to field configurations which close
along the ``space'' cycle of the torus only up to an element
$h\in\Gamma$, and then projecting onto group invariant states.
In the Lagrangian picture this corresponds to a sum over partition
functions with different boundary conditions:
For $g,h\in\Gamma$ we denote the partition function of the $h$-twisted
sector with an insertion of the operator $g$ as
\begin{equation}
  g \mathop{\Box}\limits_{\textstyle h} =
  \tr_{{\cal M}_h}\left( g
  q^{L_0-\frac{c}{24}} \bar q^{\bar L_0-\frac{c}{24}} \right) .
\end{equation}
The $\Gamma$-orbifold partition function is then obtained as
\begin{equation}
  Z[\Gamma] = \frac1{|\Gamma|}
  \sum_{\textstyle{g,h\in\Gamma\atop gh=hg}}
  g \mathop{\Box}\limits_{\textstyle h}.
\end{equation}
If $\Gamma$ is non-abelian, boundary conditions twisted by
non-commuting group elements are not consistent, hence the condition
$gh=hg$ \cite{DVaf1}.

In the case at hand we are interested in orbifolds of the symplectic
fermion model (\ref{eq:action}) by finite subgroups of $SL(2)$.  These
are the binary cyclic groups ${\cal C}_N$, the binary dihedral groups
${\cal D}_N$ and the binary tetrahedral, octahedral and icosahedral
group ${\cal T}, {\cal O}, {\cal I}$.  We will not pursue the ${\cal
  C}_N$ orbifolds for $N$ odd since they contain fermions and the
modular invariant partition function resulting from sum over spin
structures is the same as that for the ${\cal C}_{2N}$-orbifold.

Consider first the (abelian) binary cyclic group of even order ${\cal
  C}_{2N}$. We choose a torus $T$ containing the cyclic group, ${\cal
  C}_{2N} \subset T \subset SL(2)$. The generator of the torus can be
written as $K_\alpha^\beta = n_a d^{a\beta}_\alpha$, with $n_1^2 +
n_2^2 - n_0^2 = 1$, that is, the vector ${\bf n} = (n_0, n_1, n_2)$ is
a unit vector with respect to the metric $\eta$ on the group $SL(2)$
(see Appendix \ref{sec:conv}).
The cyclic group ${\cal C}_{2N}$ is then generated by $h=\exp(2\pi
iQ/N)$,
where $Q = n_a Q^a$. The fermion
field $\Phi$ has two components $\Phi^\pm$ with eigenvalues $\pm1/2$
under $Q$.
As in the $(\xi,\eta)$ system we introduce twisted sectors
${\cal M}_\lambda$, with boundary conditions
\begin{equation}
  \Phi^\pm(\e^{2\pi i}z, \e^{-2\pi i}\bar z) =
  \e^{\pm2\pi i\lambda}\Phi^\pm(z, \bar z).
\end{equation}
For the ${\cal C}_{2N}$ orbifold $\lambda$ will take the values
$\lambda=k/(2N)$ with $k=1,\ldots,2N-1$.  The boundary conditions
imply that $\Phi^\pm$ has spin $\pm\lambda$ and can be expanded in
modes as
\begin{equation}
  \Phi^\pm(z,\bar z) = i\sum_{m\in{\Bbb Z}} \left(
  \frac{\psi^\pm_{m\mp\lambda}}{m\mp\lambda} z^{-m\pm\lambda} +
  \frac{\bar\psi^\pm_{m\pm\lambda}}{m\pm\lambda}
  \bar z^{-m\mp\lambda} \right).
\end{equation}
The ground state energy in the twisted sector is as before
\begin{equation}
  h_\lambda = \bar h_\lambda = \frac{\lambda(\lambda-1)}{2}.
\end{equation}
We can again introduce the ${\cal C}_{2N}\times\mathrm{Vir}$ character
\begin{eqnarray}
  D_{\mu,\lambda}(\tau) &=&
  \tr_{{\cal M}_\lambda} \left( \e^{4\pi i\mu Q}
  q^{L_0-\frac{c}{24}} \bar q^{\bar L_0-\frac{c}{24}} \right)
  \nonumber\\*
  &=&
  (q\bar q)^{(1-6\lambda(1-\lambda))/12} \prod_{n=1}^\infty
  \left| 1 + \e^{2\pi i\mu} q^{n+\lambda-1} \right|^2
  \left| 1 + \e^{-2\pi i\mu} q^{n-\lambda} \right|^2 \\
  &=&
  |d_{\mu,\lambda}(\tau)|^2.
  \nonumber
\end{eqnarray}
which equals the character for the $(\xi,\eta)$ system. Hence the
partition function is the same as for the ${\cal C}_{2N}$ twisted
$(\xi,\eta)$ system,
\begin{equation}\label{eq:ZC}
  Z[{\cal C}_{2N}] = Z_N.
\end{equation}
To calculate the partition function $Z[\Gamma]$ for non-abelian
$\Gamma$ we follow \cite{Gins1} and add the contributions of the
mutually commuting subsets of $\Gamma$, which form cyclic groups,
subtracting any overcounting. As in \cite{Gins1} the result is
\begin{eqnarray}
  Z[{\cal D}_N] &=& \frac12(Z_N + 2Z_2 - Z_1), \\
  Z[{\cal T}] &=& \frac12(2Z_3 + Z_2 - Z_1), \\
  Z[{\cal O}] &=& \frac12(Z_4 + Z_3 + Z_2 - Z_1), \\
  Z[{\cal I}] &=& \frac12(Z_5 + Z_3 + Z_2 - Z_1).
  \label{eq:ZI}
\end{eqnarray}
Here, of course, one has to keep in mind that $Z_n(\tau)$ corresponds
to a Coulomb gas partition function at radius $n\sqrt2$ and not $n$ as
for $c=1$.

\subsection{Marginal operators}

Deformations of a conformal field theory, preserving conformal
invariance and central charge $c$, are generated by marginal
operators, that is, bosonic operators of conformal weights $(1,1)$.
In the maximal bosonic theory, the ${\cal C}_2$-model, we have eight
bosonic weight $(1,1)$ operators, four of which are obtained as
products of left times right weight one fermions,
\begin{equation}
  U(z,\bar z) = \frac12 J_{\alpha\beta}
  \psi^\alpha(z)\bar\psi^\beta(\bar z),
  \qquad
  V^a(z,\bar z) = \frac12 d^a_{\alpha\beta}
  \psi^\alpha(z)\bar\psi^\beta(\bar z).
\end{equation}
The other four are obtained by multiplying with the ``logarithmic'' field
$\omega = \frac12 J_{\alpha\beta} \Phi^\alpha\Phi^\beta$,
\begin{eqnarray}
  \tilde U(z,\bar z) &=&
  \frac14 J_{\alpha\beta} J_{\gamma\delta}
  \mathopen: \psi^\alpha(z)\bar\psi^\beta(\bar z)
  \Phi^\gamma(z,\bar z) \Phi^\delta(z,\bar z)
  \mathclose:,
  \\
  \tilde V^a(z,\bar z) &=&
  \frac14 d^a_{\alpha\beta} J_{\gamma\delta}
  \mathopen: \psi^\alpha(z)\bar\psi^\beta(\bar z)
  \Phi^\gamma(z,\bar z) \Phi^\delta(z,\bar z)
  \mathclose:.
\end{eqnarray}
The latter operators can also be obtained from the $sl(2)$ currents,
\begin{equation}
  \mathopen: J^a(z,\bar z) \bar J^b(z,\bar z) \mathclose: =
  - \frac14 f^{ab}_c \tilde V^c(z,\bar z)
  - \frac14 g^{ab} \tilde U(z,\bar z).
\end{equation}
The operators $U, \tilde U$ and $V^a, \tilde V^a$ transform as singlets
and triplets, respectively, under the charge algebra,
\begin{equation}
  \begin{array}{rcl@{\qquad}rcl}
    {}[ Q^a, U(z,\bar z)] &=& 0, &
    {}[ Q^a, V^b(z,\bar z)] &=& f^{ab}_c V^c(z,\bar z), \\
    {}[ Q^a, \tilde U(z,\bar z)] &=& 0, &
    {}[ Q^a, \tilde V^b(z,\bar z)] &=& f^{ab}_c \tilde V^c(z,\bar z).
  \end{array}
\end{equation}
These operators have vanishing 3pt functions, an integrability
condition which ensures they remain marginal when the perturbation is
switched on \cite{Kada1,KBro1}.

The effect of a perturbation by $U$ is a change the normalisation of
the action since $U$ is precisely the operator appearing in the
action,
\begin{equation}
  U = -\frac14 \partial_\mu \Phi^\alpha \partial^\mu \Phi^\beta.
\end{equation}
Under a rescaling of the action, $S \mapsto \lambda S$, the charge
algebra (\ref{eq:chargealg}) and the Virasoro algebra with central
charge $c=-2$ remain unchanged. However,
\begin{eqnarray}
  T(z) \omega(w,\bar w) &\sim&
  \frac{\lambda}{(z-w)^2} + \frac{\partial\omega(w,\bar w)}{z-w}, \\
  \bar T(\bar z) \omega(w,\bar w) &\sim&
  \frac{\lambda}{(\bar z-\bar w)^2} +
  \frac{\bar\partial\omega(w,\bar w)}{\bar z-\bar w}.
\end{eqnarray}
Perturbing with $U$ thus corresponds to change of relative
normalisation of the basis states within each $L_0$ Jordan block. It
only affects the ``logarithmic'' sector of the Virasoro symmetry and
is independent of the $SL(2)$ symmetry. In particular, perturbing an
orbifold theory $Z[\Gamma]$ by $U$ will not take us out of the
orbifold.

To consider perturbations by the marginal operators $V^a$
let us pick as before a vector ${\bf n}$ with $n_1^2 + n_2^2 -
n_0^2 = 1$.
Perturbing by $V = n_a V^a$ yields the action
\begin{eqnarray}
  S_\lambda[\Phi^\alpha] &=& S + \frac{i\lambda}{2\pi} \int\!\d^2z V
  \nonumber\\*
  &=& \int \frac{\d^2z}{2\pi} \, \left\{
  \partial\Phi^\alpha \left(J_{\alpha\beta}-\frac{i\lambda}{4}
  K_{\alpha\beta}\right) \bar\partial\Phi^\beta +
  \bar\partial\Phi^\alpha \left(J_{\alpha\beta}+\frac{i\lambda}{4}
  K_{\alpha\beta}\right) \partial\Phi^\beta \right\},
\end{eqnarray}
where $K_{\alpha\beta} = n_a d^a_{\alpha\beta}$.
The equations of motion and the propagator are the same as for the
unperturbed theory.
In fact, the perturbed action can be related to the original action by
a field redefinition.  Define new fields
\begin{equation}
  \tilde\Phi^\alpha(z,\bar z)
  = \bigl(\e^{i\Lambda K}\bigr)^\alpha_\beta \phi^\beta(z)
  + \bigl(\e^{-i\Lambda K}\bigr)^\alpha_\beta \bar\phi^\beta(\bar z),
\end{equation}
where $\Phi^\beta(z,\bar z) = \phi^\beta(z) + \bar\phi^\beta(\bar z)$.
We then obtain
\begin{equation}
  S[\tilde\Phi^\alpha] = \cos\Lambda \int \frac{\d^2z}{2\pi}
  \, \bigl\{
  \partial\Phi^\alpha \bigl(J_{\alpha\beta} +
  i\tan\Lambda K_{\alpha\beta}\bigr) \bar\partial\Phi^\beta +
  \bar\partial\Phi^\alpha \bigl(J_{\alpha\beta} -
  i\tan\Lambda K_{\alpha\beta}\bigr) \partial\Phi^\beta \bigr\}
\end{equation}
Thus, $S_\lambda[\tilde\Phi^\alpha]$ is the same as the original action
$S[\Phi^\alpha]$ with the
identification $\Lambda = -\arctan(\lambda/4)$, apart from a rescaling
by $\cos\Lambda = (\lambda^2+1/16)^{-1/2}$.
Hence, the perturbation does not change the theory and
its only effect is a change of basis from $\Phi$ to
$\tilde\Phi$ which can be accomplished by acting with the operator
$\exp(i\Lambda Q_-)$, where
\begin{equation}
  Q_- = K_{\alpha\beta} \sum_{k=1}^\infty \left(
  \frac{\psi^\alpha_{-k}\psi^\beta_k}{k} -
  \frac{\bar\psi^\alpha_{-k}\bar\psi^\beta_k}{k} \right).
\end{equation}
We expect that perturbing by $\tilde U,\tilde V^a$ again only results in
internal transformations of the theory since the operators $\tilde
U,\tilde V^a$ are the ``logarithmic'' versions of $U, V^a$.

The marginal operators in the orbifold models $Z[\Gamma]$ are those
marginal operators of the ${\cal C}_2$ model which are invariant under
$\Gamma$. Since $U$ and $\tilde U$ are $SL(2)$ singlets they survive
for any $\Gamma$. The cyclic orbifolds $Z[{\cal C}_{2N}], N>1$ contain
in addition the two operators $V = n_a V^a$ and $\tilde V = n_a \tilde
V^a$.  However, these marginal operators effect only internal
isomorphisms and thus the orbifold models are isolated
theories not linked by any $c=-2$ marginal flows.

There is a simple argument that the ${\cal C}_{2N}$ orbifolds are not
linked by a marginal flow along the circle line of Gaussian model
partition functions even though all the ${\cal C}_{2N}$ orbifold
partition functions $Z[{\cal C}_{2N}](\tau) = Z(n\sqrt2,\tau)$ lie on
that line.  For $Z(\rho,\tau)$ to describe a $c=-2$ model it needs at
least a vacuum state with $h = \bar h = 0$. This corresponds to a term
of the form $|\eta(\tau)^{-1}q^{1/8}|^2$ in the partition function
(\ref{eq:CGZ}).  Thus we need $(m\rho + n/\rho)^2 = (m\rho - n/\rho)^2
= 1/2$ for some integers $m$ and $n$ which can only be satisfied if
$\rho = n\sqrt2$ or $\rho = 1/(m\sqrt2)$.  The circle line of the
Gaussian model can, however, be interpreted for each $Z[{\cal
  C}_{2N}]$ model as a flow with the central charge varying with
radius as $c = 1 - 6N^2/\rho^2$ but leaving the effective central
charge invariant, $c_{\rm eff}=1$.  For $\rho\to1$ we flow to the
$SU(2)$ level one WZW model with a state of weight $N^2/4$ taken as
the vacuum state.

\section{Discussion}
\label{sec:disc}

We discussed two different conformal field theories at $c=-2$ --- the
$(\xi,\eta)$ system and the symplectic fermions --- and their orbifold
theories. Both theories have the same partition function and identical
primary field content and contain reducible but indecomposable
representations of the Virasoro algebra.  The ``small'' algebra is a
chiral algebra contained in both the $(\xi,\eta)$ system and the
symplectic fermions.  The fusion of twist fields yields the
``logarithmic'' operators extending the ``small'' algebra to a theory
of symplectic fermions.  These have an $SL(2)$ symmetry and orbifolds
with respect to finite subgroups of $SL(2)$ have modular invariant
partition functions reminiscent to the classification of $c=1$ modular
invariant partition functions. The partition function for the
symplectic fermion theories sees only the semi-simple part of $L_0$
since the nilpotent part drops out on taking the trace. For an attempt
to define a partition function, based on a different notion of trace,
which explicitly displays the logarithmic behaviour see \cite{Floh2}.
The orbifold models possess marginal operators which, however, only
generate internal symmetries and do not result in flows between
different orbifold models.

Fusion in conformal field theory can be formulated algebraically using
the notion of a ring-like tensor product of representations of the
chiral algebra introduced by Borcherds and developed in \cite{Gabe2},
see also \cite{Nahm1}.
We intend to give an account of structure and fusion of
``logarithmic'' representations at $c=-2$
in a forthcoming communication \cite{GKau1}.
While ``logarithmic'' fields can be studied as representations of the
Virasoro algebra using the methods presented here their interpretation
in an operator formalism and physical significance remain unclear.
The r\^ole of the ground state degeneracy and the correct renormalisation
procedure of composite operators need to be investigated further.

The significance of this study goes beyond the $c=-2$ case: For any
$c=1-6(p-p')^2/pp'$ from the minimal series, apart from the
usual minimal model, there is also a conformal field theory containing
``logarithmic'' operators. If the Virasoro highest weight
representation generated from the vacuum is irreducible the proof of
Feigin and Fuchs \cite{FeFu4} of the Virasoro fusion rules applies and
one obtains a minimal model.
But if one includes fields {\em from the edge\/} of
the Kac-table the vacuum representation becomes reducible and
the appearance of ``logarithmic'' operators
is unavoidable.
This indicates the possibility of
a wide range of two-dimensional critical phenomena where it is not
sufficient to restrict to scaling fields alone but where the product of
scaling fields yields logarithmic deviations from scaling.

The case $c=-2$ is of interest also in two-dimensional quantum gravity
since a $c=-2$ matter system coupled to gravity is exactly solvable
\cite{KKMi1,BKKM1} in terms of a matrix model \cite{Davi1,KWil1}.  The
action for the matter system is precisely the action for the
symplectic fermions presented here.  An interesting question is
whether the logarithmic scaling violations observed in the matrix
model \cite{KWil1} are related to the ``logarithmic'' operators of the
symplectic fermions.

Let us finally consider the implications for two dimensional polymers.
In \cite{Sale1} H.~Saleur argued that the dense phase of
two-dimensional polymers could be described by a $(\xi,\eta)$-system.
Specifically, the $(\xi,\eta)$-system describes the sector formed by an
even number of non-contractible polymers with modular invariant
partition function $Z_{\rm even}=Z_1$.\footnote{There is an error in
  eq. (37) of \cite{Sale1} which states $Z_{\rm even}=\frac12 Z_1$.
  However, since the partition function $Z_1$ has a unique ground
  state it can not be divided by two and it can be easily checked that
  the expression given here is correct and indeed follows from eq.
  (24) in \cite{Sale1}}
In addition there is a sector formed by an odd number of
non-contractible polymers with partition function $Z_{\rm odd}=Z_2 -
Z_1$ which corresponds to a $\Bbb Z_4$ twist of the
$(\xi,\eta)$-system.
The total polymer partition function
is the partition function $Z_2$ of the ${\cal C}_4$-orbifold of the
$(\xi,\eta)$-system.
This result was obtained by realising dense polymers as the $n\to0$
limit of the low temperature phase of the $O(n)$ model which in turn
can be mapped onto a Coulomb gas.
It agrees with the scaling dimensions
\begin{equation}\label{eq:scaldim}
  x^D_L = \frac{L^2-4}{16}.
\end{equation}
for the polymer $L$-leg operators $\Phi_L$
found by Duplantier \cite{Dupl1}.
However, the limit $n\to0$ does not commute with the thermodynamic
limit.
Furthermore, the physical quantities which have been determined for
dense polymers, the partition function $Z_2$
and the scaling dimensions (\ref{eq:scaldim}), are shared by
the ${\cal C}_4$ orbifold of both the $(\xi,\eta)$ system and the
symplectic fermions.
The Virasoro {\em primary\/} field content of both models is the same,
they differ only in the reducible Virasoro representations which are
unavoidably present.
A quantum field theory description of dense
polymers will require a new class of fields which form reducible
representations of the Virasoro algebra.
It is an interesting question whether one can see these fields in a
lattice model of polymers and whether one can then
distinguish between the twisted $(\xi,\eta)$ system and symplectic
fermions.
Numerical studies on the
lattice are likely to be difficult since one has to determine 4pt
functions in order to see these reducible Virasoro representations.
A further interesting question is whether one can see
the $SL(2)$ symmetry of the Neveu-Schwarz and Ramond
sectors in the geometrical description of the lattice model.
These considerations apply equally to the dilute phase of polymers and
percolation problems since these are described by a
non-minimal $c=0$ conformal field theory which allows ``logarithmic''
operators.

\paragraph{Acknowledgements}
I would like to thank W.~Eholzer, M.~Flohr, M.~Gaberdiel and W.~Nahm
for many useful discussions.
This work has been supported by a Research Fellowship from Sidney Sussex
College, Cambridge and partly by PPARC.
Explicit calculations were performed on computers purchased on EPSRC
grant GR/J73322 using a MAPLE programme written by the author.
I also thank the Physikalisches Institut der Universit\"at Bonn for
support and hospitality.

\appendix

\section{Theta functions}
\label{sec:theta}

The characters $d_{\mu,\lambda}$ have symmetry properties
\begin{eqnarray}
  d_{\mu+1,\lambda}(\tau) &=& \e^{-2\pi i\lambda} d_{\mu,\lambda}(\tau), \\
  d_{\mu,\lambda+1}(\tau) &=& d_{\mu,\lambda}(\tau),
  \\
  d_{-\mu,-\lambda}(\tau) &=& \e^{-2\pi i\mu} d_{\mu,\lambda}(\tau).
\end{eqnarray}
For $N$ even we can write $d_{k/N,l/N}(\tau)$ as a sum of theta
functions,
\begin{equation}
  d_{k/N,l/N}(\tau) = \eta(\tau)^{-1} \e^{-2\pi ikl/N^2}
  \sum_{n=0}^{N-1}
  \e^{2\pi ikn/N} \Theta_{Nn+l-N/2,N^2/2}(\tau),
\end{equation}
while for $N$ odd we have
\begin{eqnarray}
  d_{k/N+\epsilon/2,l/N+\epsilon'/2}(\tau) &=&
  \eta(\tau)^{-1} \e^{-\pi i(2k+N\epsilon)(2l+N\epsilon')/2N^2}
  \times\nonumber\\*&\times&
  \sum_{n=0}^{2N-1} \e^{2\pi ikn/N}
  \Biggl( \Theta_{2Nn+2l+(\epsilon'-1)N,2N^2}(\tau) +
    \nonumber\\*&&\qquad\qquad  (-1)^\epsilon
    \Theta_{2N^2+2Nn+2l+(\epsilon'-1)N,2N^2}(\tau) \Biggr),
\end{eqnarray}
where $\epsilon, \epsilon'$ are zero or one and label the different
spin structures.
Under $T\colon\tau\mapsto\tau+1$ we have
\begin{equation}
  d_{\mu,\lambda}(\tau+1) = \e^{2\pi i(\lambda(\lambda-1)/2+1/12)}
  d_{\mu+\lambda,\lambda}(\tau).
\end{equation}
For the transformation under $S\colon\tau\mapsto-\tau^{-1}$ we have
\begin{equation}
  \eta(-\tau^{-1})^{-1} \Theta_{n,m}(-\tau^{-1}) =
  \frac{1}{\sqrt{2m}} \sum_{n'=0}^{2m-1} \e^{-\pi inn'/m}
  \eta(\tau)^{-1}\Theta_{n',m}(\tau)
\end{equation}
and thus for $N$ even we obtain
\begin{equation}
  d_{k/N,l/N}(-\tau^{-1}) = \e^{\pi i/2}
  \e^{-2\pi i(k/N+1/2)(l/N)} d_{1/2-l/N,1/2+k/N}(\tau).
\end{equation}

\section{The Lie algebra $sl(2)$}
\label{sec:conv}

For the convenience of the reader we list here relations for the Lie
algebra $sl(2)$ used above.
We use Greek letters $(\alpha,\beta,\gamma,\ldots)$ for indices
referring to the fundamental representation of $sl(2)$ and letters
$(a,b,c,\ldots)$ for Lie algebra indices. All relations written in
purely tensorial notation are independent of the choice of basis.

The generators of $sl(2)$ are $2\times2$ matrices $(d^a)_\alpha^\beta$
satisfying
\begin{equation}
  {}[ d^a, d^b ] = f^{ab}_c d^c,
\end{equation}
where $f^{ab}_c$ are the structure constants of $sl(2)$ in the chosen
basis.
The Killing metric on $sl(2)$ is defined as
\begin{equation}
  g^{ab} = \tr(d^a d^b) = d_\alpha^{a\beta} d_\beta^{b\alpha}.
\end{equation}
The symplectic form $J^{\alpha\beta}$ and its inverse
$J_{\alpha\beta}$ can be used to raise and lower indices,
\begin{equation}
  d^a_{\alpha\beta} = d_\alpha^{a\gamma} J_{\gamma\beta}, \qquad
  d^{a\alpha\beta} = J^{\alpha\gamma} d_\gamma^{a\beta}
\end{equation}
and analogously for the metric $g^{ab}$ and its inverse $g_{ab}$.
The product of two representation matrices is
\begin{equation}
  d_\alpha^{a\beta} d_\beta^{b\gamma} =
  d^a_{\alpha\beta} d^{b\beta\gamma} =
  \frac12 f^{ab}_c d_\alpha^{c\gamma} +
  \frac12 g^{ab} \delta_\alpha^\gamma.
\end{equation}
Since the square of the fundamental representation decomposes into the
trivial and adjoint representation we also have
\begin{equation}
  \delta_\alpha^\gamma \delta_\beta^\delta =
  -\frac12 J_{\alpha\beta} J^{\gamma\delta}
  + d^a_{\alpha\beta} d_a^{\gamma\delta}.
\end{equation}
Contracting the above relation yields
\begin{equation}
  d_{a\alpha}^\beta d_\beta^{a\gamma} =
  \frac32 \delta_\alpha^\gamma.
\end{equation}

A convenient basis to use is the orthonormal basis
\begin{displaymath}
  d_\alpha^{0\beta} = \frac12\pmatrix{0&-1\cr1&0}, \qquad
  d_\alpha^{1\beta} = \frac12\pmatrix{0&1\cr1&0}, \qquad
  d_\alpha^{2\beta} = \frac12\pmatrix{1&0\cr0&-1},
\end{displaymath}
with commutation relations
\begin{equation}
  {}[ d^a, d^b ] = \epsilon^{abc} \eta_{cd} d^d,
\end{equation}
where $\eta_{ab} = {\rm diag}(-1,1,1)$.  The compact generator $d^0$
is anti-symmetric while the non-compact generators $d^1, d^2$ are
symmetric.  The matrices with both indices up or down read in this
basis
\begin{equation}
  d^0_{\alpha\beta} = \frac12\pmatrix{-1 & 0\cr 0 & -1}, \qquad
  d^1_{\alpha\beta} = \frac12\pmatrix{1 & 0\cr 0 & -1}, \qquad
  d^2_{\alpha\beta} = \frac12\pmatrix{0 & -1\cr -1 & 0}.
\end{equation}
An normalised $sl(2)$ element can be written as
\begin{equation}
  K_\alpha^\beta = n_i d^{i\beta}_\alpha,
\end{equation}
where ${\bf n}$ is a unit-vector with respect to the metric $\eta$,
$n_1^2 + n_2^2 - n_0^2 = 1$.
Such an element $K$ satisfies
\begin{equation}
  K_\alpha^\beta K_\beta^\gamma = \frac14 \delta_\alpha^\gamma, \qquad
  \left(\e^{i\Lambda K}\right)_\alpha^\beta =
    \cos\frac{\Lambda}{2} \delta_\alpha^\beta +
    i\sin\frac{\Lambda}{2} K_\alpha^\beta
\end{equation}
Furthermore, $K_{\alpha\beta} = K_\alpha^\gamma J_{\gamma\beta}$ is
symmetric. From this we obtain
\begin{eqnarray}
  \left(\e^{i\Lambda K}\right)_\alpha^\gamma J_{\gamma\delta}
    \left(\e^{-i\Lambda K}\right)_\beta^\delta =
  \cos\Lambda J_{\alpha\beta} + i\sin\Lambda K_{\alpha\beta} .
\end{eqnarray}

\end{document}